\documentclass[aps,prl,reprint,amsmath,amssymb,notitlepage,citeautoscript,superscriptaddress]{revtex4-2}

\usepackage{bm,mathrsfs,dcolumn,graphicx,color}
\usepackage{amsmath}
\usepackage{graphicx}
\usepackage[T1]{fontenc}
\usepackage{hyphenat}
\usepackage{setspace}

\usepackage{siunitx} % \si{\angstrom}

\usepackage[normalem]{ulem}
\definecolor{darkerblue}{rgb}{0,0,0.75}
\definecolor{darkerred}{rgb}{0.8,0,0}
\usepackage[colorlinks=true,citecolor=darkerblue,linkcolor=darkerred]{hyperref}
\newcommand{\exciting}{{\usefont{T1}{lmtt}{b}{n}exciting}}

\begin{document}

\title{Robust excitons across the phase transition of two-dimensional hybrid perovskites}

\author{Jonas D. Ziegler}
\affiliation{Dresden Integrated Center for Applied Physics and Photonic Materials and Würzburg-Dresden Cluster of Excellence ct.qmat, Technische Universität Dresden, 01062 Dresden, Germany}
\affiliation{Department of Physics, University of Regensburg, 93053 Regensburg, Germany}
\author{Kai-Qiang Lin}
\affiliation{Department of Physics, University of Regensburg, 93053 Regensburg, Germany}
\author{Barbara Meisinger}
\affiliation{Department of Physics, University of Regensburg, 93053 Regensburg, Germany}
\author{Xiangzhou Zhu}
\affiliation{Department of Physics, Technical University of Munich, 85748 Garching, Germany}
\author{Manuel Kober-Czerny}
\affiliation{Department of Physics, University of Oxford, Clarendon Laboratory, Oxford OX1 3PU, United Kingdom}
\author{Pabitra K. Nayak}
\affiliation{Department of Physics, University of Oxford, Clarendon Laboratory, Oxford OX1 3PU, United Kingdom}
\affiliation{TIFR Centre for Interdisciplinary Sciences, Tata Institute of Fundamental Research, 36/P, Gopanpally Village, Serilingampally Mandal, Hyderabad 500046, India}
\author{Cecilia Vona}
\affiliation{Institut f\"ur Physik and IRIS Adlershof, Humboldt-Universit\"at zu Berlin, Berlin Germany}
\author{Takashi Taniguchi}
\affiliation{International Center for Materials Nanoarchitectonics,  National Institute for Materials Science, Tsukuba, Ibaraki 305-004, Japan}
\author{Kenji Watanabe}
\affiliation{Research Center for Functional Materials, National Institute for Materials Science, Tsukuba, Ibaraki 305-004, Japan}
\author{Claudia Draxl}
\affiliation{Institut f\"ur Physik and IRIS Adlershof, Humboldt-Universit\"at zu Berlin, Berlin Germany}
\author{Henry J. Snaith}
\affiliation{Department of Physics, University of Oxford, Clarendon Laboratory, Oxford OX1 3PU, United Kingdom}
\author{John M. Lupton}
\affiliation{Department of Physics, University of Regensburg, 93053 Regensburg, Germany}
\author{David A. Egger}
\affiliation{Department of Physics, Technical University of Munich, 85748 Garching, Germany}
\author{Alexey Chernikov}
\email{alexey.chernikov@tu-dresden.de}
\affiliation{Dresden Integrated Center for Applied Physics and Photonic Materials and Würzburg-Dresden Cluster of Excellence ct.qmat, Technische Universität Dresden, 01062 Dresden, Germany}
\affiliation{Department of Physics, University of Regensburg, 93053 Regensburg, Germany}

\begin{abstract}
Two-dimensional halide perovskites are among intensely studied materials platforms profiting from solution based growth and chemical flexibility.
They feature exceptionally strong interactions among electronic, optical as well as vibrational excitations and hold a great potential for future optoelectronic applications.
A key feature for these materials is the occurrence of structural phase transitions that can impact their functional properties, including the electronic band gap and optical response dominated by excitons.
However, to what extent the phase-transitions in two-dimensional perovskites alter the fundamental exciton properties remains barely explored so far.
Here, we study the influence of the phase transition on both exciton binding energy and exciton diffusion, demonstrating their robust nature across the phase transition.
These findings are unexpected in view of the associated substantial changes of the free carrier masses, strongly contrast broadly considered effective mass and drift-diffusion transport mechanisms, highlighting the unusual nature of excitons in two-dimensional perovskites.
\\
\\
\end{abstract}
\maketitle

Coulomb-bound electron-hole pairs, commonly known as excitons, dominate the optical response of nanostructured materials\,\cite{Haug2009}.
A prominent example are two-dimensional (2D) Ruddlesden-Popper halide perovskites hosting tightly-bound, mobile exciton quasiparticles\,\cite{Ishihara1989,Tanaka2005,Gauthron2010,Yaffe2015,Blancon2018,Baranowski2019,Cho2019,Cho2021}. 
Excitons in 2D perovskites are primarily responsible for the absorption and emission of light, governing key material properties for high-efficiency photovoltaic and lighting applications\,\cite{Kumar2016,Tsai2016,Wang2016b,Tsai2018,Chen2018,Liang2021,Chen2020}.
They combine very high binding energies up to 0.5\,eV with sizable mobilities\,\cite{Deng2020,Seitz2020,Xiao2020}, exhibiting efficient transport from ambient conditions down to cryogenic temperatures\,\cite{Ziegler2020}.
These properties render the material system a promising platform for future optoelectronics and offers a rich playground to study fundamental physics of interacting quasiparticles.

In addition to strongly bound excitons, 2D halide perovskites incorporate a soft crystal lattice featuring highly efficient exciton-phonon coupling and possible polaronic effects \,\cite{Gauthron2010,Straus2016,Guo2016,Wright2016,Urban2020,SrimathKandada2020}.
A particularly interesting phenomenon in this regard are the structural phase transitions\,\cite{Kind1980,Billing2007} that 2D perovskites inherit from their bulk counterparts.
It involves temperature-induced changes of the crystal symmetry with more recent findings indicating a dynamic order-disorder transition\,\cite{Menahem2021}.
Phase transitions in 2D perovskites often occur close to room-temperature, strongly affect the optical properties\,\cite{Ishihara1990,Martin-Garcia2021}, are sensitive to the layer thickness\,\cite{Yaffe2015}, and were recently shown to change the effective masses of the charge carriers\,\cite{Baranowski2019, Dyksik2020}.
They represent a key feature of this class of materials with potential implications for the design of perovskite-based optoelectronic devices and their response to temperature.

However, the fundamental properties of excitons in 2D perovskites remain essentially unexplored at the phase transition.
While a recent study finds strong changes in the diamagnetic shifts\,\cite{Baranowski2019} attributed to changes of the exciton total mass, the impact of the phase transition on both binding energy and diffusion of excitons are not known.
Moreover, the occurrence of a structural phase transition offers a unique opportunity to investigate fundamental properties of excitons including their transport behavior in a well-defined experimental scenario.
This is particularly relevant for the case of 2D halide perovskites, as the excitons represent a considerable challenge for appropriate descriptions, exacerbated by their intermediate Wannier-Frenkel nature and the strong coupling to phonons.  

Here, we address these questions by studying excitonic properties across a structural phase transition in 2D perovskites.
Using a combination of non-linear spectroscopy and transient microscopy we determine temperature-dependent exciton binding energies and monitor exciton diffusion at temperatures directly above and below the phase transition.
First-principles calculations demonstrate changes of the carrier masses between the two phases that are concurrent with modification of the single-particle band gap accounting for the observed spectral shift.
Surprisingly, both exciton binding energies and the diffusion coefficients are found to be constant between low- and high-temperature phases. 
These results strongly suggest that excitonic properties do not trivially scale with free carrier masses, challenging the appropriate description of excitons and their dynamics in 2D halide perovskites.

In our study we used thin layers exfoliated from 2D butylammonium-lead-iodide crystals, $(n=1)$ BA$_2$-PbI$_4$, with about 50\,nm thickness, encapsulated between hBN layers for environmental protection\,\cite{Seitz2019} and transferred onto SiO$_2$/Si substrates.
The samples were placed into helium-cooled microscopy cryostats for optical measurements. 
We used a 80\,MHz tunable 140\,fs Ti:Sapphire laser for excitation in photoluminescence (PL) experiments and a tungsten-halogen whitelight source for reflectance.
The incident light was focused to a spot with about 1\,$\mu$m diameter.
The PL was then either dispersed by a grating or deflected by a mirror to obtain spectrally- and spatially-resolved response, respectively.
A CCD-sensor was used for time-integrated signals and we employed a streak-camera for time-resolved detection; additional details are provided in the Supplementary Information (SI).

\begin{figure}[t]
	\centering
			\includegraphics[width=8.5 cm]{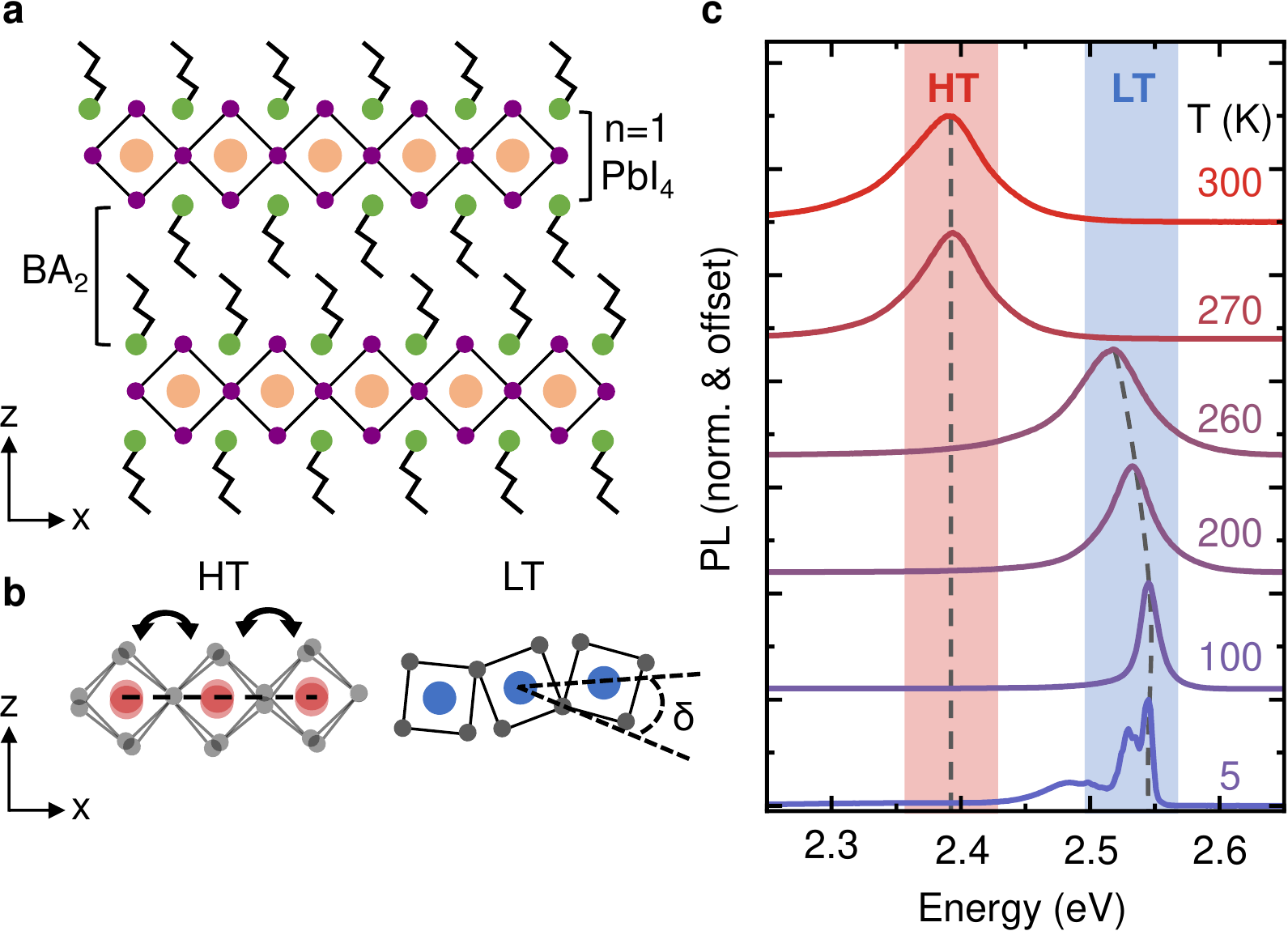}
		\caption{\textbf{Structural phase transition in two-dimensional perovskites}
		\textbf{a.} Schematic illustration of the layered structure of $(n=1)$ BA$_2$PbI$_4$, consisting of alternating organic (butylammonium, BA) and inorganic layers (PbI$_4$) with layer number $n$. 
		\textbf{b.} Illustration of thermally-induced anharmonic fluctuations of the octahedral tilting in the high-temperature phase (HT), and an effective locking with a larger average distortion $\delta$ in the low temperature phase (LT). 
		\textbf{c.} Measured PL spectra of an hBN-encapsulated BA$_2$PbI$_4$ thin layer for different temperatures across the phase transition at T$_{PT} = 270 K$.
		}
	\label{fig1}
\end{figure}

\textbf{Phase Transition in 2D Perovskites.}
The structure of the studied BA$_2$PbI$_4$ crystals is schematically illustrated in Fig.\,\ref{fig1}\,(a). The inorganic layers composed of PbI$_4$ octahedra host the electronic states forming conduction and valence bands at the fundamental band gap of the material. The organic chains between the layers effectively serve as barriers in this natural multi-quantum-well structure. 
The phase transition, driven by the ordering of the butylammonium chains, changes the relative alignment of the octahedra between the phases\,\cite{Ishihara1990}. In the high-temperature phase (HT), recent findings emphasize both strong anharmonic fluctuations of the tilt angle and indications of an order-disorder transition\,\cite{Menahem2021}. In the low-temperature (LT) phase, the octahedra become effectively locked at a larger average distortion angle $\delta$ \,\cite{Dyksik2020} illustrated in Fig.\,\ref{fig1}\,(b).

The impact of the phase transition on the optical response of layered alkylammonium lead-iodide perovskites is primarily associated with a spectral shift on the order of 100\,meV\,\cite{Ishihara1990, Yaffe2015}. PL spectra of the studied BA$_2$PbI$_4$ samples presented in Fig.\,\ref{fig1}\,(c) demonstrate this characteristic effect. As the temperature increases, the dominant exciton peak abruptly shifts to lower energies by 120\,meV at about $T=270$\,K, corresponding to the known temperature range of the phase transition in this class of materials\,\cite{Billing2007, Ishihara1990}. In addition, the spectra exhibit a continuous, smooth increase of the spectral linewidth from thermal population of vibrational modes and increasingly rapid exciton-phonon scattering at higher temperatures\,\cite{Gauthron2010}.

\begin{figure}[t]
	\centering
		\includegraphics[width=6.5 cm]{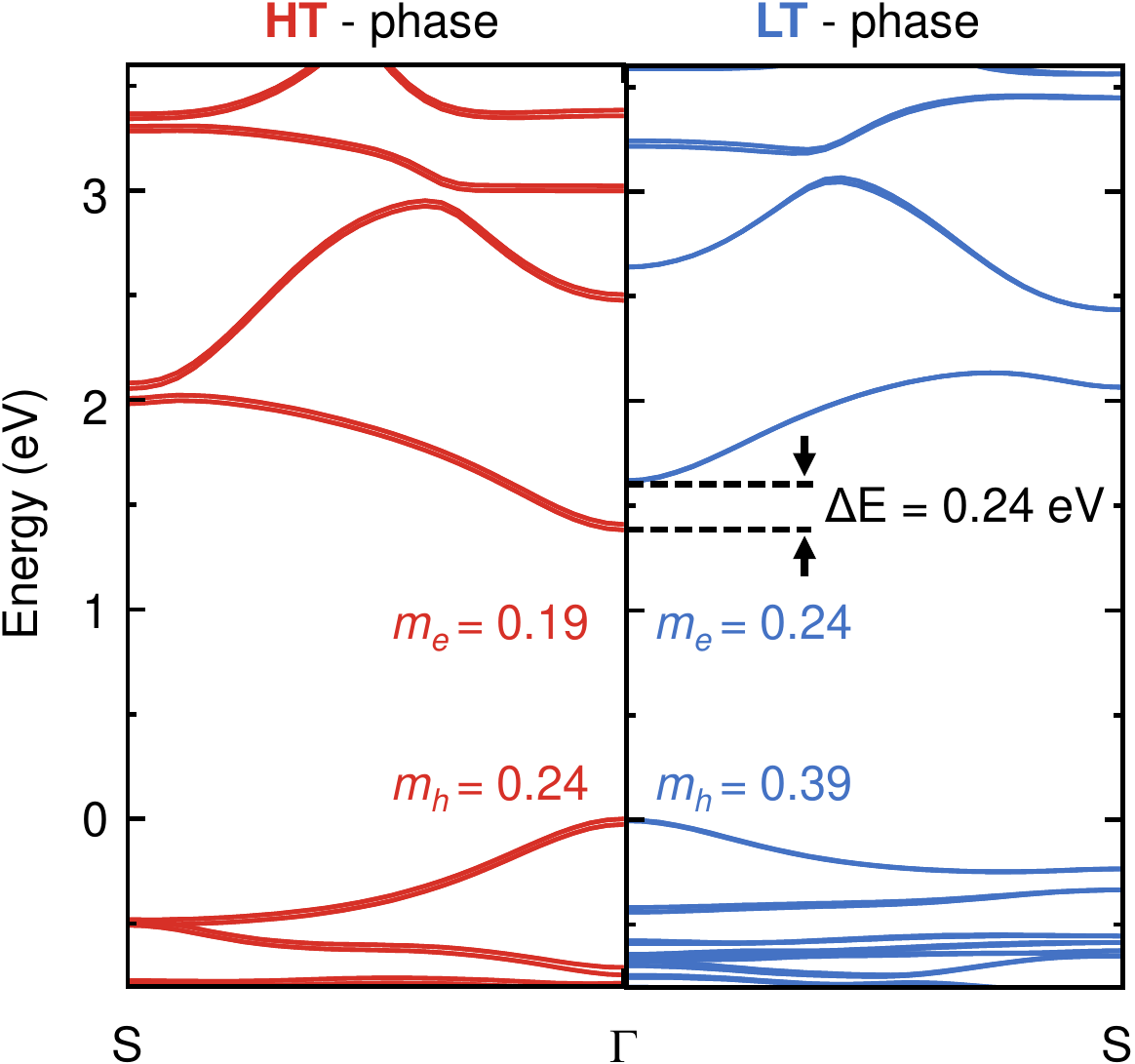}
		\caption{\textbf{Electronic band structure across the phase transition.} The Kohn-Sham band energies, computed by PBE including SOC are shown between $\Gamma$ and S (S=0.5 0.5 0) for the optimized crystal structures of BA$_2$PbI$_4$. 
		The VBM of both phases are aligned at zero. 
		The trends in the differences of the band gap and effective masses are consistent between PBE and HSE06 (see SI and Ref. \cite{Vona2022_inpreparation}).
			}
	\label{fig2}
\end{figure}

For a theoretical analysis, we perform electronic-structure calculations using density-functional theory (DFT). 
To this end we utilize two different DFT codes: VASP \,\cite{Kresse1996}, implementing the projector augmented wave potentials \cite{Kresse1999} uses a plane-wave basis set, and the all-electron full-potential code \exciting\ \cite{Gulans2014}, based on the linearized augmented planewave method. 
We first apply the PBE functional\,\cite{Perdew1996}, augmented by van-der Waals corrections calculated with the Tkatchenko-Scheffler method\,\cite{Tkatchenko2009} to describe exchange-correlation interactions, and include the effects of the spin-orbit coupling (SOC). 
Following previous work \,\cite{Dyksik2020}, in Fig.\,\ref{fig2} we show the main changes of the electronic band structure across the phase transition. 
We observe a substantial difference of 240\,meV in the electronic band gap between the HT and LT phase. 
Thereby SOC has a substantial contribution as verified by additional calculations for a model structure where the BA cations are replaced by Cs (see SI and Ref. \cite{Vona2022_inpreparation}). 
More specifically, without accounting for SOC we find that the gap changes by only about 110\,meV going from the LT to the HT phase. 

Using the same procedure, we also perform calculations with the hybrid functional HSE06 \cite{Krukau2006} as implemented in \texttt{exciting} \cite{Vona2022}. 
As expected, the HSE06 functional has a strong impact on the band-gap, increasing it in both phases by $\sim$0.6 eV (see Fig.\,\ref{fig2}) but, hence, does not alter the differences across the phase transition. 
Similarly, the absolute values of the effective masses that are substantially lower in the HT phase are confirmed by both PBE and HSE06 calculations.

\begin{figure*}[ht]
	\centering
			\includegraphics[width=14 cm]{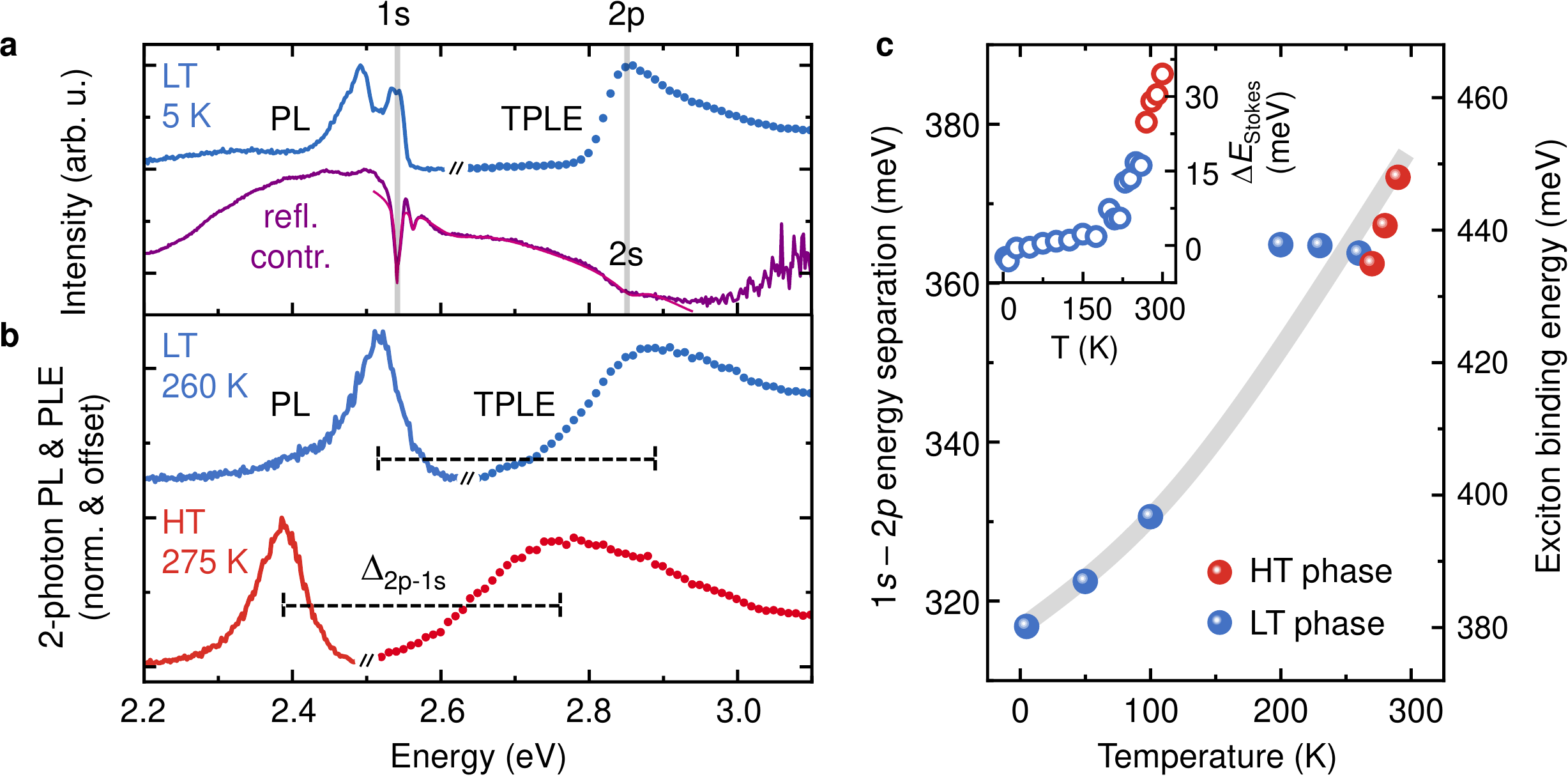}
		\caption{\textbf{Temperature-dependent exciton binding energy }
		\textbf{a.} Two-photon photoluminescence excitation (TPLE) and PL spectra of BA$_2$PbI$_4$ at $T=5$\,K.
		The reflectance contrast spectrum corresponding to the changes of reflectance relative to the substrate is presented in the same graph for comparison. 
		\textbf{b.} PL and TPLE spectra measured for the LT and HT phases at $T=260$\,K and $T=275$\,K, respectively. 
		The spectra are corrected for changes in the effective absorption due to the multi-layer interference effects. 
		The energy separation between ground and excited state $\Delta_{2p-1s}$, proportional to the respective exciton binding energy, is indicated by the dashed lines. 
		\textbf{c.} Temperature-dependent $\Delta_{2p-1s}$ and the corresponding exciton binding energy $E_b$, indicated on the right y-axis. 
	For this analysis, the $1s$ energy was determined from reflectance contrast to exclude influence of the temperature-dependent Stokes shift, presented in the inset. 
	The fitting error is below the size of spheres and the gray line is a guide to the eye.
		}
	\label{fig3}
\end{figure*}

Hence, based on various electronic-structure methods, we find that part of the experimentally observed energy shift, can  already be attributed to effects that occur at the single-particle level.
Specifically, the effect is a direct consequence of the changes in the average out-of-plane distortion angle of PbI$_4$ octahedra \,\cite{Dyksik2020}:
a smaller distortion angle in the HT phase leads to a stronger electronic coupling between individual octahedra and hence to a larger bandwidth and a smaller band gap.
For the same reason, the effective masses are substantially smaller in the HT than in the LT phase, which is in line with previous measurements of the diamagnetic shifts and theoretical work \,\cite{Baranowski2019}. 
Moreover, by lifting certain degeneracies of the Pb states SOC manifests itself in additional changes of the band gap and effective masses. 
Therefore, when assuming Wannier-type excitons and band-like transport, one would expect both exciton binding energy and exciton diffusion to change accordingly across the phase transition.

\begin{figure*}[ht]
	\centering
			\includegraphics[width=13 cm]{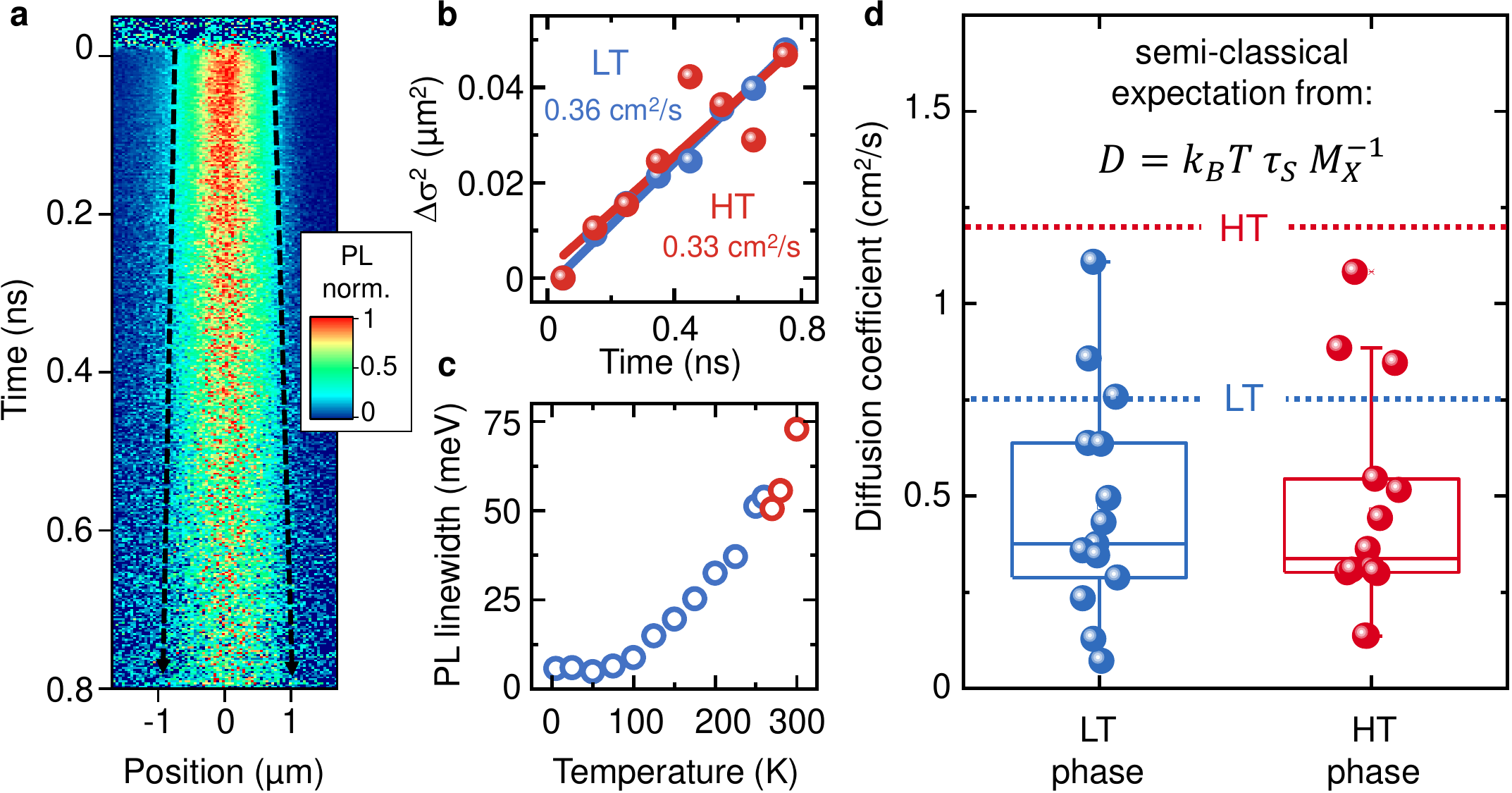}
		\caption{\textbf{Exciton diffusion at the phase transition} 
		\textbf{a.} Representative streak camera image of the spatially- and time-resolved exciton emission for the LT phase ($T = 260$\,K). 
		The data is normalized at each time step. 
		\textbf{b.} Extracted change of the mean-squared displacement $\Delta\sigma^2$ for the LT and HT phases as a function of time. 
		Diffusion coefficients are obtained from the slope $\Delta\sigma^2(t)$ indicated by solid lines. 
			\textbf{c.}  Temperature dependent PL linewidth determined from Lorentz fits presented as full-width-at-half-maximum, corresponding to the change of the exciton-phonon scattering rate.
		\textbf{d.} Box chart summarizing measured exciton diffusion coefficients ($N=29$) across the phase transition for both phases in the temperature range between 260 and 274\,K. 
		The spread in the data stems from different samples and sample positions. Estimated values from a semi-classical drift-diffusion model based on calculated effective masses and scattering times obtained from PL linewidths are indicated.	
		}
	\label{fig4}
\end{figure*} 

\textbf{Exciton binding energy.}
To measure exciton binding energy as a function of temperature across the phase transition we employed two-photon photoluminescence excitation (TPLE) spectroscopy; details of the setup are given in the SI and Ref. \cite{Lin2021}.
The PL signal of the ground state exciton  above 2.3 eV was detected as a function of the laser photon energy tuned between 1.24 and 1.77\,eV.
In this spectral range, exciton states with $p$-type orbital symmetry determine two-photon absorption\,\cite{Wang2005} and the resulting TPLE signal is usually dominated by the $2p$ transition\,\cite{Wang2005,Ye2014,Berkelbach2015}.
The advantage of TPLE, in contrast to linear spectroscopy, is the suppression of direct optical transitions allowing us to detect spectrally broad, weaker signals from excited exciton states both at low and elevated temperatures.
An exemplary TPLE spectrum of a BA$_2$PbI$_4$ sample is presented in the top panel of Fig.\,\ref{fig3}\,(a) together with the corresponding PL data at lattice temperature of 5\,K.
Both the ground and excited states are in good agreement with the measured reflectance contrast spectrum, shown for direct comparison. 
The analysis of the reflectance spectra using a transfer matrix approach further allows us to correct the measured TPLE spectra for effective two-photon absorption from multi-layer interferences (see SI).

The strongest resonance in the reflectance spectra and the high-energy peak of the PL at 2.54\,eV stems from the ground-state exciton transition ($1s$).
It features weaker replica on the high energy side, commonly attributed to phonon-sidebands\,\cite{Straus2016,Urban2020} mimicking the low-energy side of the PL at 5\,K (c.f. Fig.\,\ref{fig1}\,(b)).
The maximum of the TPLE spectrum at 2.85\,eV corresponds to the $2p$ energy with the high-energy flank attributed to weaker contributions of $3p$, $4p$,... etc. resonances.
In addition, the $2p$ peak closely corresponds to the energy of the $2s$ transition weakly visible in the reflectance spectrum.
The energy separation between $2p$ and $1s$ as well as the near-degeneracy of $2p$ and $2s$ excited states are consistent with large exciton binding energies and the properties of excited states in 2D perovskites\,\cite{Tanaka2005,Yaffe2015,Blancon2018,Cho2019}.

Figure \ref{fig3}\,(b) presents PL and TPLE spectra for two temperatures close to the structural phase transition in the HT and LT regime.
At these elevated temperatures, the $1s$ and $2p$ resonances broaden and their energy separation slightly increases in comparison to the 5\,K data.
The main difference between HT and LT spectra is a shift of both ground and excited state peaks to lower energies of about 120$\pm10$\,meV.
The magnitude of the shift is thus essentially the same as for the single-photon PL presented in Fig.\,\ref{fig1}\,(c), compared to values between 110 and 240\,meV predicted by first-principles calculations (c.f. Fig.\,\ref{fig2}).
Their relative separation $\Delta_{2p-1s}$, however, remains constant.
Temperature-dependent values for $\Delta_{2p-1s}$ are summarized in Fig.\,\ref{fig3}\,(c).
Here, the $1s$ energy is determined from reflectance contrast instead of PL to exclude the influence of the temperature-dependent Stokes shift, presented in the inset. 
The Stokes shift is on the order of 20\,meV for both LT and HT regions at the phase transition and changes smoothly as a function of temperature.
The exciton binding energy $E_b$ is then estimated by taking advantage of linear scaling with the energy separation $\Delta_{2p-1s}$.
We use the relationship $E_b = 1.2\times\Delta_{2p-1s}$ from the literature\,\cite{Cho2019} and present temperature-dependent binding energy on the right y-axis in Fig.\,\ref{fig3}\,(c). 

The absolute values for the exciton binding energy around 400\,meV in BA$_2$PbI$_4$ correspond well to the previously reported low-temperature results\,\cite{Tanaka2005,Yaffe2015,Blancon2018}.
Interestingly, however, the binding energy does not change between LT and HT phases across the phase transition.
This result is rather surprising, since the reduced effective mass $\mu=m_hm_e/(m_{h}+m_e)$ (see Fig.\,\ref{fig2} (b)) changes from 0.15 (LT) to 0.11\,m$_0$ (HT) across the phase transition. 
Since the exciton binding energy typically scales linearly with the reduced mass in the Wannier model\,\cite{Haug2009}, these changes should lead to a sharp \textit{decrease} of the binding energy by as much as about 40\% in the HT phase.
Instead, the exciton binding energy is not only constant between LT and HT phases, but also appears to continuously increase with temperature up to 20\% from 5 to 290\,K.
We note that if we neglect the influence of the Stokes shift, the increase of the binding energy would appear even more pronounced. 

\textbf{Exciton Diffusion Coefficient.}
Following the analysis of exciton binding energies, we examine the exciton diffusion -- a process commonly expected to be affected by the \textit{total} mass of the exciton.
Experimentally, exciton propagation is monitored by detecting spatially- and time-resolved PL for a series of temperatures across the phase transition.
The samples are excited using energy densities from a few to several 10's of nJ/cm$^2$ per pulse.
We confirm the absence of density-dependent effects such as exciton-exciton annihilation\,\cite{Warren2000} (see SI) and perform a series of measurements on different samples and spot positions.
A typical streak camera image of the time-dependent cross-section $x$ is shown in Fig.\,\ref{fig4}\,(a) for the LT phase at $T=260$\,K.
The signal is normalized at each time step, illustrating continuous broadening of the exciton cloud.
For the analysis the PL intensity profiles are fitted with Gaussian distributions $\propto exp[-x^2/2\sigma(t)^2]$ at time intervals of $\Delta t=100$\,ps.
The extracted time-dependent change of the variance $\Delta\sigma(t)^2$ is presented in Fig.\,\ref{fig4}\,(b) for LT and HT phases at the temperatures of 260 and 270\,K, respectively.
The corresponding phase is verified ahead of the diffusion measurements at every temperature and sample position using spectrally resolved PL, as shown in Fig.\,\ref{fig1}\,(c).

As it is characteristic for diffusive propagation, the variance increases linearly with time for both LT and HT cases.
In contrast to the spectral shift of the emission energy, however, the diffusion coefficients $D$ extracted from the slope of $\Delta\sigma^2=2Dt$\,\cite{Ginsberg2020} are essentially the same.
This observation is accompanied by a nearly continuous increase of the exciton-phonon scattering rate with temperature, reflected in the change of the PL spectral linewidth\,\cite{Gauthron2010} in Fig.\,\ref{fig4}\,(c).
Within the framework of the semi-classical band-like propagation model, the diffusion coefficient scales with the temperature $T$, exciton mass $M_X$ and scattering rate $\hbar/\tau$ as $D=(k_BT\times\tau)/M_X$.
Consequently, we would expect the diffusivity to follow the change of the total exciton mass by a factor of 1.5 across the phase transition (see Fig.\,\ref{fig2}), taking into account the nearly constant scattering rate.  
In contrast to that, the nearly equal diffusion coefficients of the LT and HT phase mimic the lack of change in the exciton binding energy.
We further note that the absolute values of the diffusion coefficients vary for different flakes and sample positions, as illustrated by the box chart in Fig.\,\ref{fig4}\,(d).
Nevertheless, we do not find any systematic differences of the diffusion coefficients across the phase transition, compared to the semi-classical expectation values from calculated exciton masses (c.f. Fig.\,\ref{fig2}).
 
\textbf{Discussion}
Excitons in 2D perovskites combine seemingly conflicting properties: on the one hand, they show large exciton binding energies that are typically associated with localized Frenkel-type states. 
On the other, the relatively low effective masses and the efficient exciton transport suggest wavefunctions extending over many unit cells, which is typically a Wannier-type feature.
The latter consideration also manifests itself in the successful application of effective mass models with modified thin-film Coulomb potentials in combination with first-principles calculations to describe key exciton properties such as binding energies and Bohr radii\,\cite{Tanaka2005,Blancon2018,Baranowski2019,Cho2019,Cho2021}.
These models typically involve parameters such as appropriate dielectric screening constants of the organic and inorganic layers and the reduced exciton mass.

Our study provides a unique opportunity to assess such models for BA$_2$PbI$_4$, since the phase transition is associated with a change in the \textit{average} crystal structure, as mentioned above.
Indeed we find, in line with previous work, that along with the change of the structure the band gap and effective masses of the system are also modified across the phase transition.
Interestingly, while the band gap change is at least qualitatively reflected in the spectral shift of the optical transition, the difference of effective masses does not manifest itself in any accompanying changes of the exciton binding energy and diffusion constant.
Clearly, the canonical effective-mass theory for these two key properties does not appear fully consistent with the experimental findings.
Consequently, our understanding of the exciton properties of 2D perovskites should be reconsidered, potentially along the following lines.

In this context, the observation of a constant exciton binding energy at the structural phase transition strongly implies that the dependence on the effective mass of the \textit{free} carriers is substantially weaker than commonly assumed.
It is conceivable that the exciton wavefunction could spread over a sufficiently large region of the reciprocal space beyond the effective mass approximation, where the effect of the phase transition on the band curvature may be reduced.
Another possibility involves the change of the dielectric screening\,\cite{Guo2019} in the appropriate frequency range, given by the binding energy, that offsets the difference of the carrier masses.
It is also possible that the relevant values for the quasiparticle masses are determined by polaronic effects that may not change too drastically at the phase transition. 
Renormalization of the effective mass could also be connected to the observed increase of the exciton binding energy with rising temperature.

Similar considerations can be suggested for exciton diffusion in BA$_2$PbI$_4$.
In 3D halide perovskites, a dynamic disorder picture has been put forward, which takes into account the strong \textit{local fluctuations in the lattice} away from the average structure and rationalizes key transport properties in this manner\,\cite{Mayers2018,Lacroix2020,Schilcher2021}.
While there are important differences between 3D and 2D perovskites, it is interesting that anharmonic fluctuations were recently demonstrated also for the 2D systems\,\cite{Menahem2021}.
Since exciton diffusion appears to be essentially immune to changes in the effective mass of the average structure, one could speculate that it could be also largely governed by local fluctuations of the lattice.
It is still intriguing, however, that the semi-classical description seems to account for temperature-dependent transport of free excitons in a related system, 2D phenylethylammonium-lead iodide\,\cite{Ziegler2020}.
However, the general applicability of such models can be challenged in view of the exciton mean free path being typically smaller than their wavepacket size, given by the de Broglie wavelength\,\cite{Glazov2020, Schilcher2021}. 
The observed inadequacy of the semi-classical prediction for the change of the diffusion coefficients at the phase transition provides additional experimental evidence that alternative descriptions for the exciton transport in 2D perovskites should be explored.

\subsection*{Conclusions}
In summary, we have explored the impact of the structural phase transition in 2D perovskites on the exciton properties.
Temperature-dependent exciton binding energy and exciton diffusion was monitored across the phase transition using a combination of non-linear spectroscopy and time-resolved microscopy.
Both exciton binding energy and exciton diffusion are found to be unaffected by the phase transition.
These findings go beyond the expectations of the effective mass description of the excitons and the semi-classical understanding of their transport.
Our results highlight the unusual behavior of excitons in this class of materials and strongly motivate further developments for their description.
Moreover, the robust nature of the excitons with respect to substantial changes of the crystal structure seems particularly promising from the perspective of applications.

\subsection*{Acknowledgements}
We thank our colleagues Matan Menahem, Omer Yaffe, Paulina Plochocka, and Michal Baranowski for helpful discussions.
Financial support by the DFG via SPP2196 Priority Program (Project-ID: 424709454) and Emmy Noether Initiative (CH 1672/1, Project-ID: 287022282) as well as the W\"urzburg-Dresden Cluster of Excellence on Complexity and Topology in Quantum Matter ct.qmat (EXC 2147, Project-ID 390858490) is gratefully acknowledged. 
D.A.E. acknowledges funding from the Alexander von Humboldt Foundation within the framework of the Sofja Kovalevskaja Award, endowed by the German Federal Ministry of Education and Research; the Technical University of Munich - Institute for Advanced Study, funded by the German Excellence Initiative and the European Union Seventh Framework Programme under Grant Agreement No. 291763; the DFG under Germany's Excellence Strategy - EXC 2089/1–390776260, and the Gauss Centre for Supercomputing e.V. (www.gauss-centre.eu) for funding this project by providing computing time through the John von Neumann Institute for Computing (NIC) on the GCS Supercomputer JUWELS at Jülich Supercomputing Centre (JSC).
K.W. and T.T. acknowledge support from the Elemental Strategy Initiative conducted by the MEXT, Japan (Grant Number JPMXP0112101001) and  JSPS KAKENHI (Grant Numbers 19H05790, 20H00354 and 21H05233). 
C.V. acknowledges the North-German Supercomputing Alliance (HLRN) for providing computational resources.

%%%%%%%%%%%%%%%%%%  REFERENCES %%%%%%%%%%%%%%%%%%%%%%%%%%%

%\bibliography{references}

%%%%%%%%%%%%%%%%%%%%%%%%%%%%%%%%%%%%%%%%%%%%%%%%%%%%%%%%%%


\end{document}